\documentclass[twocolumn,preprintnumbers,amsmath,amssymb]{revtex4-2}
\usepackage{graphicx}
\usepackage{dcolumn}
\usepackage{balance}
\usepackage{ulem}
\usepackage[svgnames]{xcolor}
\usepackage{bm}
\usepackage{multirow}
\usepackage{amsmath}
\usepackage{gensymb}
\makeatletter
\newcommand*{\balancecolsandclearpage}{%
  \close@column@grid
  \clearpage
  \twocolumngrid
}
\makeatother

\begin{document}

\title{Phonon-Mediated Superconductivity near the Lattice Instability in Hole-doped Hydrogenated Monolayer Hexagonal Boron Nitride} 
\author{Takat B. Rawal$^{1,*}$, Ling-Hua Chang$^{1}$, Hao-Dong Liu$^{2}$, Hong-Yan Lu$^{2, *}$, and C. S. Ting$^{1}$} 

\affiliation{$^{1}$Texas Center for Superconductivity and Department of Physics, University of Houston, Houston, TX 77204, USA\\
$^{2}$School of Physics and Physical Engineering, Qufu Normal University, Qufu 273165, China}
\date{\today}

\begin{abstract}
Employing the density-functional theory with local density approximation, we show that the fully hydrogenated monolayer-hexagonal boron nitride (H$_2$BN) has a direct-band gap of 2.96 eV in the blue-light region while the pristine \textit {h}-BN has a wider indirect-band gap of 4.78 eV. The hole-doped H$_2$BN is stable at low carrier density ($n$) but becomes dynamically unstable at higher $n$. We predict that it is a phonon-mediated superconductor with a transition temperature ($T_c$) which can reach $\sim$31 K at $n$ of $1.5\times$ $10^{14}$ holes cm$^{-2}$ near the lattice instability. The $T_c$ could be enhanced up to $\sim$82 K by applying a biaxial tensile strain at 6 \% along with doping at $n$ of $3.4\times$ 10$^{14}$ holes cm$^{-2}$ close to a new lattice instability. 
\end{abstract}

\maketitle

Two-dimensional (2D) electron system such as graphene has attracted a lot of attention since its atomic structure in the form of a single layer was realized experimentally \cite{Novoselov1}. Owing to the exceptional electronic and other properties, it may be used for a wide range of applications \cite{Novoselov2, Geim, Morozov, Neto, Kotov, Li, JoelI-Jan}, including superconducting quantum circuits for quantum computing \cite{JoelI-Jan}. By tuning its electronic properties, one may remarkably achieve and adjust the superconductivity. Very recently, the robust superconductivity has been observed for the magic angle twisted trilayer graphene \cite{Park1} with $T_c$ up to 2.1 K \cite{Hao}. For the twisted bilayer graphene, the $T_c$ up to 1.7 K has been reported \cite{Cao1,Cao2}. The superconductivity may arise in twisted trilayer or bilayer graphene from interlayer interactions which are missing in its monolayer limit, albeit the full understanding of the mechanism behind the superconductivity in a graphene-like, 2D system is not well developed.\\
\indent The study of another 2D system like the monolayer-hexagonal boron nitride (\textit {h}-BN) begins to become popular recently. The pristine compound has a honeycomb lattice structure similar to that of graphene. It has been predicted \cite{Topsakal} that the monolayer \textit {h}-BN is an insulator with an indirect band gap of 4.47 eV, but recent experiments \cite{Park} demonstrate that it has a gap of $\sim$5 eV. By employing the first-principles methods in prior studies on 2D materials \cite{Profeta, Savini, Ludbrook, Lu, YLu}, the doped \textit {h}-BN has been predicted to be a phonon-mediated superconductor with $T_c$ $\sim$41 K under an applied biaxial tensile strain together with doped holes ($5.3\times$ 10$^{14}$ holes cm$^{-2}$)\cite{Jin}. The doped holes may come from either gating or chemical doping. The gating approach is similar to the case for biaxial-tensile-strained and doped graphene ($4.0\times$ 10$^{14}$ holes cm$^{-2}$) which shows the superconductivity with $T_c$ $\sim$30 K \cite{Si}. By means of chemical doping of Ca, Ba, and Sr atoms, one may theoretically obtain the superconductivity in monolayer \textit {h}-BN with $T_c$ ranging from $\sim$1 K to 10 K \cite{ShimadaA}. The bilayer \textit {h}-BN with an intercalated Li atom has also been predicted to show the superconductivity with different $T_c$ ($<$ 25 K) \cite {ShimadaB, Szewczyk}. Despite some efforts being devoted to developing the fundamental understanding of phonon-mediated superconductivity in doped \textit {h}-BN, the electronic structure, lattice stability, and superconductivity in H$_2$BN (fully hydrogenated monolayer \textit {h}-BN) have not been studied. The main focus of this letter is to investigate these properties in pristine and hole-doped H$_2$BN by adopting an approach similar to the one implemented in previous studies on graphane \cite{Savini} and graphene \cite{Si}.\\
\indent In this letter, using methods based on the first-principles density functional theory (DFT) with local density approximation (LDA) plus plane waves and norm-conserving pseudopotentials approaches \cite{Perdew, Kohanoff} and the density functional perturbation theory (DFPT) \cite{Baroni}, we show that hole-doped H$_2$BN is an electron-phonon superconductor. We show the DFT-optimized structure of H$_2$BN in Fig.1. Firstly, we study the electronic band structures of pristine \textit {h}-BN and H$_2$BN (see Figs.S1a and S1b) in the supplemental material \cite{AA}. We found that the undoped H$_2$BN is an insulator with a direct-band gap of 2.96 eV in the blue-light region. The full hydrogenation of \textit {h}-BN gives rise to a transition from indirect \cite{Topsakal} to the direct-band gap. The subsequent modification in electronic properties by hole doping gives the metallicity with the finite electron density of states at the Fermi level. We show that there exist no negative phonon frequencies in k-space for the doped H$_2$BN with the carrier density $n$ varying from 0 up to $1.5\times$ 10$^{14}$ holes cm$^{-2}$, indicating the stability of the lattice in this doping range. But as $n>\sim1.55\times$ 10$^{14}$ holes cm$^{-2}$, negative phonon frequencies begin to appear near the $\Gamma$ point implying that the lattice becomes unstable \cite{AA}. For the stable lattice structures, we evaluate the electron-phonon (e-ph) interaction strength $\lambda$ in hole-doped H$_2$BN using DFPT \cite{Baroni}. Taking advantage of the high-phonon frequencies and the soft phonon modes near the lattice instability, we examine the existence of the phonon-mediated superconductivity in hole-doped H$_2$BN and estimate its $T_c$. Our methods are based upon the Bardeen-Cooper-Schrieffer (BCS) theory \cite{Bardeen} and the Eliashberg approach plus its extended versions \cite {Eliashberg, Allen, Marsiglio, McMillan}. The modified Eliashberg theory \cite{McMillan, Allen} has been widely employed to understanding the BCS superconductivity in metallic materials. We show that with hole doping at $n=1.5\times$ 10$^{14}$ holes cm$^{-2}$, the doped H$_2$BN exhibits phonon-mediated superconductivity with $T_c$ which can reach $\sim$31 K above the boiling point of liquid hydrogen. The $T_c$ can be enhanced further above the liquid nitrogen temperature when we apply biaxial tensile strain (BTS) at 6 \% and increase the hole-carrier density up to $3.4\times$ 10$^{14}$ holes cm$^{-2}$ near the new lattice instability.\\ 
\begin{figure}[h!]
  \includegraphics[scale=1]{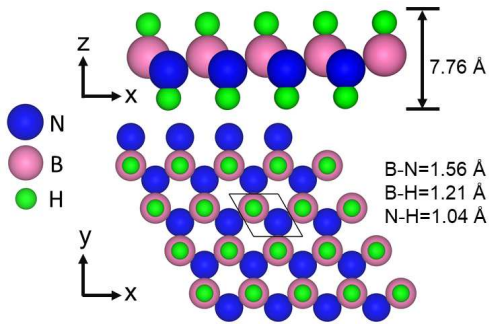}
  \caption{DFT-optimized structure of a H$_2$BN monolayer: side (top panel) and top views (bottom). The thickness of the H$_2$BN is 7.76 \AA. The calculated bond lengths are also depicted. The ($1\times1$) unit cell of the H$_2$BN is also shown by black line (bottom).}
  \label{fig:1}
\end{figure}\\
\indent The hydrogenation of \textit {h}-BN modifies its electronic structure along with band gap \cite{AA} owing to the charge redistributions that trigger electrons to be piled up at different potential regions from those of monolayer pristine \textit {h}-BN. The valence band dispersion does not follow the periodicity of the Brillouin zone (BZ) of the \textit {h}-BN and it is due to the different periodic potential of H$_2$BN with the nearly 1/2 symmetry operations (excludes the 60° inverse and 180° rotational symmetries) from that of the pristine monolayer \textit {h}-BN. Without hydrogenation, the LDA indirect-band gap turns out to 4.78 eV (see Fig.S1a). The calculated indirect-band gap of monolayer \textit{h}-BN at the level of DFT generalized-gradient approximation is 4.47 eV \cite{Topsakal}, whereas the most recent experiment reveals the gap of $\sim$5 eV \cite{Park}. Our result is thus in better accord with the experiment. The band gap of monolayer \textit{h}-BN is found to be smaller than the bulk band gap of $\sim$6.0 eV \cite{Cassabois}, where the quantum confinement of electrons is more dominant than that of a monolayer. The LDA band gap, however, reduces to 2.96 eV when monolayer \textit {h}-BN is fully hydrogenated (Fig.S1b). Both valence band maximum and conduction band minimum occur at the center of BZ, and therefore there is a cross-over from indirect to direct band gap transition. Thus, we can tune the band gap of the monolayer \textit {h}-BN from indirect to direct with a value changed by 1.82 eV. Such electronic modification may be useful for exploiting H$_2$BN in optical quantum devices.\\
\indent Fig.2a shows the electronic band structure of hole-doped H$_2$BN with the carrier density of $1.0\times$ 10$^{14}$ holes cm$^{-2}$ along the high-symmetry lines K-$\Gamma$-M-K and also its total electron density of states (DOS) (This hole concentration can be obtained by replacing 0.06 \% of N atoms with the same amount of B atoms, and such hole-doped compound can be represented by H$_2$B$_{(1+x)}$N$_{(1-x)}$ with x=0.006. The chemical potential labeled by 0 eV crosses the top of the valence bands near the zone center of BZ. Thus, it creates two hole pockets: one with a lighter mass and the other with a heavier mass. With the doping of holes increased to $1.5\times$ 10$^{14}$ holes cm$^{-2}$, the chemical potential moves downward toward the lower energy and crosses the top of the valence band near the high-symmetry K point. The Fermi surfaces at these doping levels are respectively shown in Figs.2b and 2c. For these doping levels, the Fermi surfaces are found to appear near the $\Gamma$ point. The hole pockets are essentially originated from the 2$p_x$ and 2$p_y$ orbitals of B and N atoms. Because of the two dimensionalities in the electronic band structure, the $\sigma$ bands contribute strongly to the DOS at the Fermi level \cite{MAn}. The occupied $\sigma$-bonding bands should promote remarkably the e-ph coupling (EPC) in H$_2$BN, as is the case for MgB$_2$ \cite{Bekaert, MAn} and graphane \cite{Savini}. 
\begin{figure}[h!]
  \includegraphics[scale=1]{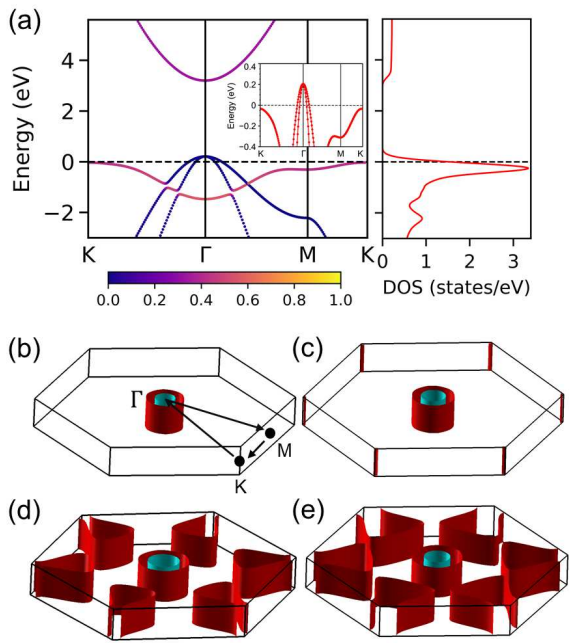}
    \caption{(a) Electronic band structure along high-symmetry directions in the first Brillouin zone (left) and the electronic density of states (right) of the hole-doped H$_2$BN with the carrier density ($n$) of $1.0\times$ 10$^{14}$ holes cm$^{-2}$. The Fermi level $E_F$ is set to zero. The color bar represents the projections of H s states to bands. For clarity of the effect of hole doping, the band structure for the small energy window around the $E_F$ is also shown in inset. Total DOS is shown by the solid line. (b-c) Fermi surfaces in the first Brillouin zone of the doped H$_2$BN with $n$ of (b) $1.0\times10^{14}$ holes cm$^{-2}$ and (c) $1.5\times$ 10$^{14}$ holes cm$^{-2}$. In (b), the high-symmetry k-point directions (K-$\Gamma$-M-K) are also depicted by arrow. By applying biaxial-tensile strain of 6 \%, the Fermi surfaces are also shown: (d) $2.0\times10^{14}$ holes cm$^{-2}$ and (e) $3.4\times10^{14}$ holes cm$^{-2}$.} 
   \label{fig:2}
\end{figure}\\
\indent We examine the orbital contributions to the top valence band of hole-doped H$_2$BN. Figs.S2a and S2b show the projected electronic DOS for pristine \textit {h}-BN and H$_2$BN, respectively. The projected DOS of the doped H$_2$BN with $1.0\times$ 10$^{14}$ holes cm$^{-2}$ is shown in Fig.S3. The top valence bands at the $\Gamma$ point (Fig.2a) are dominant with N 2$p_x$ and 2$p_y$ orbital characters and have minor contributions from B 2$p_z$ and H 1s orbitals, thus suggesting the top valence bands are $\sigma$-like bonding bands. The projected DOS due to the B 2$p_z$ and H1s orbitals are almost identical in the energy range from approximately $-1$ to 0.2 eV (Fig.S3). In the same energy range, the bands at high-symmetry K-points originate dominantly from the B 2$p_z$ and H1s orbitals.\\
\indent We now determine the vibrational properties of hole-doped H$_2$BN using DFPT \cite{Baroni}. Fig.3 shows the resulting phonon band structure and phonon DOS of hole-doped H$_2$BN at the carrier density of $1.0\times$ 10$^{14}$ holes cm$^{-2}$. The phonon dispersion of doped H$_2$BN exhibits the kinks at $\bm {q}<2\bm {k}_F$ (Fig.3) where a discontinuity results in the momentum dependence of the e-ph interaction. Such effect coincides with the Kohn anomaly \cite{Kohn}. The calculated phonon spectra without imaginary frequencies substantiate the dynamical stability of the doped H$_2$BN model. As shown in Fig.S4, the phonon modes softening occurs at and around the $\Gamma$ point that is similar to the doped fully hydrogenated graphene \cite{Savini}. There exist six different optical modes, as shown in Fig.S5. The optical modes with frequencies at 3321 cm$^{-1}$, 2328 cm$^{-1}$, and 846 cm$^{-1}$ (see Figs.S5a, S5b, and S5e) have polarizations nearly perpendicular to the lattice plane at the $\Gamma$ point. Our numerical analysis indicates that frequencies related to these modes are little affected by doping (see Fig.3 and Fig.S6a). With respect to the pristine H$_2$BN, the hole-doped H$_2$BN has the largest optical B-N stretching mode softening \cite{AA} that takes place at the $\Gamma$ point with the frequency changing from 850 to 675 cm$^{-1}$ for a carrier density $1.0\times$ 10$^{14}$ holes cm$^{-2}$ (see Fig. S4), and from 850 to 647 cm$^{-1}$ for $1.5\times$ 10$^{14}$ holes cm$^{-2}$. In Table S1, we compare the optical phonon frequencies with the degenerate modes at the $\Gamma$ point of BZ of doped H$_2$BN at the carrier density up to $1.5\times$ 10$^{14}$ holes cm$^{-2}$. These optical phonon frequencies become softer when we increase the carrier density. From Fig.S4, we can see that the acoustic modes become soft in a doped sample. The H-related in-plane motion gives rise to three different frequencies with higher phonon DOS around 1000 cm$^{-1}$ in Fig.3. The softer optical phonon modes with frequencies near 1059 cm$^{-1}$ give rise to the highest phonon DOS. In addition, there are two other DOS peaks, one at $\sim$1205 cm$^{-1}$ and the other at $\sim$864 cm$^{-1}$. Although the phonon DOS is relatively small near 965 $^{-1}$, the optical phonon frequencies are related to the in-plane motion of H atoms (Fig.S4 and S5) and thus should effectively contribute to the e-ph interaction. The contribution of these modes to the e-ph interaction has been theoretically studied previously \cite{Giustino} and the numerical calculations here are based on this approach. With the increase of carrier density up to $1.5 \times$ 10$^{14}$ holes cm$^{-2}$, the lattice is still dynamically stable, as shown in Fig.S6a. But if we increase the carrier density to $1.6\times$ 10$^{14}$ holes cm$^{-2}$, the lattice becomes unstable with negative frequencies appearing in one of the acoustic modes near the zone center (Fig.S6b and inset). Here, we study the phonon-mediated superconductivity up to the hole density close to the lattice instability.\\ 
\indent The onset of the lattice instability appears to occur at $\sim1.55\times$ 10$^{14}$ holes cm$^{-2}$ under no BTS. It may arise from the fact that the Fermi level is trying to cross the top of the valence bands at K-points and to create the hole pockets. Near K-points the hole bands are dominantly originated from B 2$p_z$ and H 1s orbitals which contribute a high DOS peak as shown in Fig.S3.\\ 
\indent By applying BTS \cite{Jin, Si}, defined by $\epsilon$=($a-a_0$)/$a_0$ $\times$ 100\% where $a$ and $a_0$ are the in-plane lattice constants for strained and unstrained ($a_0$=2.56 \AA) cases respectively, we can push the lattice instability to the higher hole density. For instance, if $\epsilon=$6 \% is applied, the doped system with hole density of $3.4\times$ 10$^{14}$ holes cm$^{-2}$ that is unstable at $\epsilon=0$, now becomes stable (see Fig.S7a). But when we increase the hole density to $3.5\times$ 10$^{14}$ holes cm$^{-2}$, the lattice again exhibits instability (see Fig.S7b and inset). Under BTS of 6 \%, we plot the Fermi surfaces in Fig.2d and 2e, respectively, for $2.0\times$ 10$^{14}$ holes cm$^{-2}$ and $3.4\times$ 10$^{14}$ holes cm$^{-2}$. As shown, the Fermi surfaces contain hole pockets near both the high-symmetry $\Gamma$ and K-points.
\begin{figure}[h!]
  \includegraphics[scale=1]{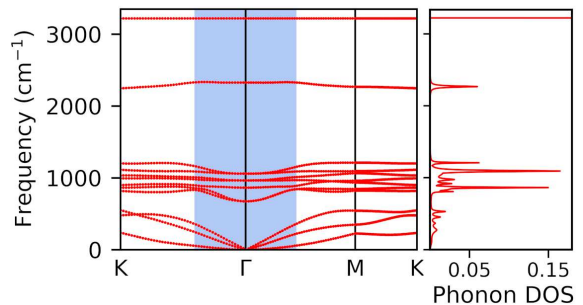}
  \caption{Phonon frequency dispersion along high-symmetry directions of the first Brillouin Zone (left) and phonon density of states (right) of the hole-doped H$_2$BN with the carrier density at $1.0\times$ 10$^{14}$ holes cm$^{-2}$. The modes with $\bm {q}<2\bm {k}_F$ are shaded.}
  \label{fig:3}
\end{figure}\\
\indent Using equation (1), we compute phonon linewidths ($\gamma_{\bm q\nu}$) which we use later for evaluating the Eliashberg function.
\begin{eqnarray}
\gamma_{\bm q\nu}=2\pi\omega_{\bm q\nu}\sum_{mn} \int_{BZ} \frac{d\bm k}{A_{BZ}} &|g_{mn}^\nu(\bm {k,q})|^2 \delta(E_{\bm k,m}-E_F)\nonumber\\ &\times\delta(E_{\bm{k+q},n}-E_F)
\end{eqnarray}
where $g_{mn}^\nu(\bm {k,q}$) is the e-ph matrix elements for an electron with momentum $\bm {k}$ and band index \textit{m} and \textit{n}, and for a phonon with wave vector $\bm {q}$, branch index $\nu$, and frequency $\omega_{\bm q\nu}$, $A_{BZ}$ the BZ area, $\delta$ the Dirac delta, and $E_F$ Fermi energy. Since holes at the top of the zone-centered, $\sigma$-bonding valence-band couple strongly to the optical bond-stretching modes \cite{Boeri}, the phonons corresponding to soft optical B-N stretching modes may strongly couple with charge carriers in the hole-doped H$_2$BN.\\ 
\indent The connection between the e-ph matrix elements and the EPC is established by the Eliashberg relation \cite{Grimvall}:
\begin{equation}
    \alpha^{2}F(\omega)=\frac{1}{2\pi N(E_F)}\sum_{\bm q\nu} \delta{(\omega-\omega_{\bm q\nu})} \frac{\gamma_{\bm q\nu}}{ \omega_{\bm q\nu}}
\end{equation}
where $N(E_F)$  is the electronic DOS at $E_F$, $\gamma_{\bm q\nu}$ (equation 1) is the phonon linewidth of mode $\nu$ with phonon momentum $\bm {q}$ and frequency $\omega_{\bm {q}\nu}$, and $\delta$ the Dirac delta. Fig.4 shows the Eliashberg spectral function of doped H$_2$BN that measures the relative contribution of its different modes to the superconducting pairing \cite{Grimvall}. For the doping with $1.0\times$ 10$^{14}$ holes cm$^{-2}$, the Eliashberg functions possess a maximum peak at 756 cm$^{-1}$ (Fig.4(a)), closer to the $\Gamma$-point frequency of 675 cm$^{-1}$ (Fig.S5), that originates from the in-plane B-N stretching modes. In the optical region, the other two peaks appear at 978 cm$^{-1}$ and 1079 cm$^{-1}$ which are associated with the H in-plane vibrations. On the circular Fermi surface near the $\Gamma$ point, electrons strongly couple with phonons corresponding to the in-plane B-N modes and moderately with those corresponding to the in-plane vibrations of H atoms. For the higher carrier density at $1.5\times$ 10$^{14}$ holes cm$^{-2}$, a maximum peak appears at 629 cm$^{-1}$ (Fig.4(b)), closer to the $\Gamma$-point frequency of 647 cm$^{-1}$, that is associated with the in-plane B-N stretching modes, and the other two peaks at 966 cm$^{-1}$ and 1076 cm$^{-1}$ which are associated with the in-plane optical modes of H atoms. 
\begin{figure}[h!]
  \includegraphics[scale=1]{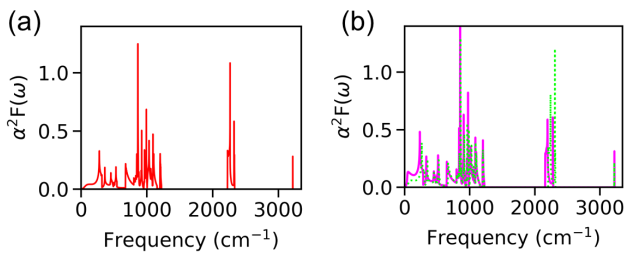}
  \caption{Eliashberg function of hole-doped H$_2$BN with the carrier density of a) $1.0\times$ 10$^{14}$  holes cm$^{-2}$ and b) $1.3\times$ 10$^{14}$ holes cm$^{-2}$ (green trace) and $1.5\times$ 10$^{14}$ holes cm$^{-2}$ (magenta).}
  \label{fig:4}
\end{figure}\\
\indent Based on the calculated Eliashberg function and vibrational properties, we expect that the phonon modes with the maximal coupling to the charge carriers would be the B-N in-plane stretching modes with H atoms oscillating in phase with the B and N atoms. The modes with moderate coupling would be the in-plane optical modes associated with H atoms. The former modes are similar to those in the p-doped graphane that has the in-plane C-C stretching and H atoms moving in phase with C atoms for both 1\% and 4\% dopings \cite{Savini}.\\
\indent We calculate the superconducting transition temperature using the modified McMillan equation \cite{Allen}: 
\begin{equation}
    T_c= \frac{<\omega_{log}>}{1.2}exp \left [\frac{-1.04(1+\lambda)}{\lambda-\mu^{*}(1+0.62\lambda)} \right]
\end{equation}
where $\mu^{*}$ is the Coulomb parameter (or Morel-Anderson Coulomb pseudopotential), the $\lambda$ is given by $2\int d\omega\omega^{-1}\alpha^{2}F(\omega)$ \cite{Grimvall}, and $<\omega_{log}>$ is the logarithmically averaged phonon frequency and is given by $exp[2\lambda^{-1}\int d\omega\omega^{-1}\alpha^{2}F(\omega)log\omega]$, with $\alpha^{2}F(\omega)$ being the Eliashberg spectral function, as given by equation 2. In atomic units, the reduced Planck's constant ($\hbar$) and Boltzmann constant ($k_B$) can be set to unity.\\
\setlength{\tabcolsep}{10pt}
\begin{table}[h!]
\caption{Hole carrier density (\textit{p} in holes cm$^{-2}$), biaxial tensile strain ([$\epsilon$] in percentage), electronic density of states at $E_F$ (EDOS in states/eV), e-ph coupling parameter ($\lambda$), logarithmically averaged frequency ($<\omega_{log}>$ in cm$^{-1}$), superconducting transition temperature (T$_c$ in Kelvin) for hole-doped H$_2$BN.}
\begin{tabular}{c c c c c c} 
\hline
\textit{p} ($\times 10^{14}$) [$\epsilon$] &   EDOS     & $\lambda$     & $<\omega_{log}>$ & T$_c$ \\  [0.5ex]
\hline
0.6 [0]    &   0.42     &  0.61       &  580    &  14.3 \\  [0.5ex]
 
1.0 [0]    &    0.61    &  0.72       &  499    &  20.7 \\  [0.5ex]
 
1.3 [0]    &    0.69    &  0.81       &  428    &  24.0 \\  [0.5ex]
  
1.5 [0]    &    0.86    &  1.10       &  312    &  31.3 \\  [0.5ex]
 
1.6 [6]    &    1.00    &  1.20       &  398    &  45.5 \\  [0.5ex]

2.0 [6]    &    1.06    &  1.67       &  386    &  64.2 \\  [0.5ex]

2.8 [6]    &    1.11    &  2.25       &  365    &  76.5 \\  [0.5ex]
 
3.4 [6]    &    1.32    &  2.75       &  350    &  82.5 \\  [0.5ex]
\hline
\end{tabular}
\end{table}\\
 \indent In Table I, we present the calculated superconducting transition temperature and the EPC $\lambda$ of the doped H$_2$BN for different densities of hole carriers. The $T_c$ appears to be increased from 14.3 K at the carrier density (\textit{n}) of $0.6\times$ 10$^{14}$ holes cm$^{-2}$ to 31.3 K at \textit{n}$=1.5\times$ 10$^{14}$ holes cm$^{-2}$. Within this range of hole density, both the electron DOS and $\lambda$ get enhanced with the increase of $n$. This result shows the correlation between EPC, DOS, and robust superconductivity \cite{Choi}. In our calculations for $T_c$, we adopted $\mu^{*}=0.13$ which was used in \cite{Savini} (note also that values of $\mu^{*}$ are ranged from 0.10 to 0.14 \cite{Jin, ShimadaA, ShimadaB}). It is important to notice that doping with higher carrier densities would cause the lattice to become unstable at $1.6\times$ 10$^{14}$ holes cm$^{-2}$ (see Fig.S6b). The critical hole density for creating the instability is $n_c\sim1.55\times$ 10$^{14}$ holes cm$^{-2}$. But we can stabilize the lattice instability by applying BTS. If we take $\epsilon=$6 \%, the hole density that causes the lattice instability is moved to higher density $n_c\sim 3.45\times$ 10$^{14}$ holes cm$^{-2}$ (see Fig.S7). In Table I, we also list the calculated $T_c$ for doped H$_2$BN under BTS of 6 \% with several $n$ ranged from 1.6 to 3.4 $\times$ 10$^{14}$ holes cm$^{-2}$. When $n$ approaches the value near the lattice instability, the DOS, EPC, and $T_c$ increase. We estimate the $T_c$ as high as 82.5 K above the liquid nitrogen temperature at $n=3.4\times$ 10$^{14}$ holes cm$^{-2}$. The $T_c$ might be further increased by applying more BTS along with the higher doping, however, the physics regarding the phonon-mediated superconductivity in hole-doped H$_2$BN will not be changed.\\
 \indent To examine the spin-polarized effect in hole-doped H$_2$BN, we perform calculations up to $1.5\times$ 10$^{14}$ holes cm$^{-2}$ at $\epsilon$=0 and find that there are negligible magnetizations and the magnetic moment on each atom, i.e., less than 0.01 $\mu_B$ atom. Therefore, the electronic states of H$_2$BN  within this doping limit are non-magnetic. We also perform similar calculations for the doped H$_2$BN under BTS at $\epsilon=$6 \% and with carrier density up to $3.4\times$ 10$^{14}$ holes cm$^{-2}$ and find no magnetic states.\\ 
\indent In summary, we have demonstrated using first-principles DFT approaches that the p-doped H$_2$BN can be an e-ph superconductor. The chemical modification of monolayer \textit {h}-BN by hydrogen atoms results in not only high-energy phonons but also the higher phonon DOS. This full hydrogenation process also leads to the transition from the indirect-wider to a direct-narrower band gap. The favorable doping of H$_2$BN subsequently by holes can form a stable lattice with strong e-ph couplings. The synergy between hydrogenation and doping plays a crucial role for superconductivity in hole-doped H$_2$BN. The phonon-mediated superconductivity with $T_c$ of 31.3 K can be achieved for the doped H$_2$BN with $n=1.5\times$ 10$^{14}$ holes cm$^{-2}$ near the lattice instability when $\epsilon$=0. To understand superconductivity at higher $n$, we need to apply a finite BTS for pushing a critical hole density toward the higher $n$. At $\epsilon=6$\%, we can dope the H$_2$BN with the higher $n$ up to $3.4\times$ 10$^{14}$ holes cm$^{-2}$ close to a new lattice instability, and thereby enhance $T_c$ up to 82.5 K. We can thus achieve the maximum $T_c$ at hole densities close to lattice instabilities. This is quite different from the case of doped \textit {h}-BN, where no superconductivity is found at $\epsilon=0$, but $T_c$ is estimated to be 41 K at $\epsilon$=17.5 \% and $n=5.3\times$ 10$^{14}$ holes cm$^{-2}$ \cite{Jin} with $\mu^{*}=0.10$. We applied weaker BTS and chose smaller hole densities with the expectation that they may be more accessible to experiments. Since the fully hydrogenated graphene has been realized in experiments \cite{Son}, we expect that there should be no obstacles to grow the fully hydrogenated monolayer \textit {h}-BN in the laboratory. We hope that predicted superconductivity in doped H$_2$BN will be confirmed experimentally in the future. When realized, one may exploit the doped H$_2$BN for nano-superconducting quantum devices that may have potential applications in quantum information technology.\\ 
\begin{acknowledgments}
This work was supported by the Texas Center for Superconductivity at University of Houston,  the Robert A. Welch Foundation (Grant No. E-1146), and the National Natural Science Foundation of China (Grant No. 12074213). We acknowledge the Research Computing Data Core (RCDC) at the University of Houston for providing the high-performance computing resources. First-principles calculations were performed using the Quantum Espresso \cite{Giannozzi} packages.\\
\end{acknowledgments}

\noindent * \verb|tbrawal@gmail.com|\\
* \verb|hylu@qfnu.edu.cn|

\balancecolsandclearpage
\onecolumngrid
\renewcommand{\thesection}{S\arabic{section}}
\renewcommand{\thefigure}{S\arabic{figure}}
\setcounter{section}{0}
\setcounter{figure}{0}
\begin{center}
{\large \bf Supplemental Material for "Phonon-Mediated Superconductivity near the Lattice Instability in Hole-doped Hydrogenated Monolayer Hexagonal Boron Nitride"}
\end{center}
\section {Computational methods}
\textbf {Density functional theory}. We perform first-principles density functional theory (DFT) calculations using plane waves and pseudopotentials approaches \cite{Perdew, Kohanoff}. Our DFT calculations utilize the local density approximation (LDA) functional \cite {Perdew} implemented within the Quantum Expresso (QE) code \cite {Giannozzi}. The interaction between ions and electrons are treated using norm-conserving pseudopotentials \cite {Hamann, Kresse}, which take into account B-2s$^2$,2p$^1$, N-2s$^2$,2p$^3$, and H-1s$^1$ as valence electrons. We use the LDA for describing the exchange-correlation of electrons since it reproduces very well the geometrical structure of \textit{h}-BN \cite {Kern, Hamdi}. To achieve the convergence of the total energy below 0.07 mRy per atom, we use an energy cutoff of 80 Ry for the expansion of wave functions of valence electrons in plane waves. For the electronic integration during self-consistent cycles, we describe the fractional occupations using the first-order Methfessel-Paxton method \cite {Methfessel}. To simulate the two-dimensional hexagonal boron nitride (\textit{h}-BN) system, we use ($1\times1$) unit cell containing 16 $\AA$ of vacuum which is found sufficient to prevent artificial electrostatic interaction between layers along z direction. We initially place hydrogen atoms alternatively on both sides of the \textit{h}-BN plane. Here, we denote the corresponding geometry of the fully hydrogenated monolayer \textit{h}-BN as H$_2$BN. For structural relaxations, we use the $24\times24\times1$ k-mesh with automatic generation of k-points according to the Monkhorst-Pack scheme \cite {Monkhorst}. The positions of atoms are optimized using the Broyden–Fletcher–Goldfarb–Shanno (BFGS) quasi-newton algorithm so that forces on each atom reach below 10$^{-7}$ Ry/a.u.. The optimized lattice parameter of the fully hydrogenated ($1\times1$) \textit{h}-BN is 2.56 \AA, which is deviated by $\sim$2\% from that of crystalline \textit{h}-BN (2.51 \AA) \cite {Solozhenko}. In the rigid-band approximation \cite {Noffsinger}, we simulate the hole doping into H$_2$BN through replacing the total valence electrons per N atom per unit cell ($Z_B+Z_N+2Z_H=3+5+2=10$e) by ($10-$y)e where y represents the number of holes per unit cell. The similar method has been adopted in prior studies on p-doped graphane \cite {Savini}.\\
\\
\indent \textbf {Density functional perturbation theory}. To compute phonon modes and dispersions as well as electron-phonon (e-ph) interaction, we use density functional perturbation theory (DFPT) \cite {Baroni} implemented in QE \cite {Giannozzi}. The total number of perturbations due to atomic displacements to be treated amounts to 3N, where N=Number of atoms, i.e. 12 for the ($1\times1$) H$_2$BN monolayer. For both the self-consistent and non self-consistent electronic structure calculations, we again use the kinetic energy cutoff of 80 Ry for plane wave basis and the cutoff of 488 Ry for the augmentation charges. We calculate the phonon frequencies using the linear-response technique within DFPT. For calculations of e-ph interaction, we adopt a scheme of interpolation over the Brillouin Zone. We compute the e-ph coupling using the momentum k-mesh of $192\times192\times1$. Such dense k-points mesh is found to sufficient for calculating the electronic density of states. For calculation of phonon density of states, we use a dense $110\times110\times1$ mesh of q-points with ndos value of 2500. We checked in details the convergence of e-ph interaction parameter ($\lambda$) for several values of smearing for both electrons and phonons. We calculate the superconducting critical temperature using the modified McMillan equation \cite {Allen} and Coulomb parameter $\mu^{*}$=0.13 \cite {Allen, Grimvall}. The choice of other values of $\mu^{*}$ do not affect qualitatively our conclusions.
\section {Electronic structure of pristine and hydrogenated monolayer \textit{h}-BN}
Fig.S1 shows the calculated electronic band structures of monolayer \textit {h}-BN (pristine) and fully hydrogenated monolayer \textit {h}-BN (H$_2$BN) along the high-symmetry directions in the first Brillouin zone. The Valence band (VB) maximum occurs at K point for \textit{h}-BN whereas it occurs at $\Gamma$ point for H$_2$BN. For both cases, the conduction band (CB) minimum occurs at $\Gamma$. Within the local density approximation, the indirect-band gap for monolayer \textit{h}-BN (pristine) is found to 4.78 eV and the direct-band gap for undoped H$_2$BN is 2.96 eV. Our calculated band structure for monolayer \textit {h}-BN is in consistent with the prior DFT calculations \cite {Topsakal}.\\ 
\begin{figure*}[h!]
\includegraphics[scale=1]{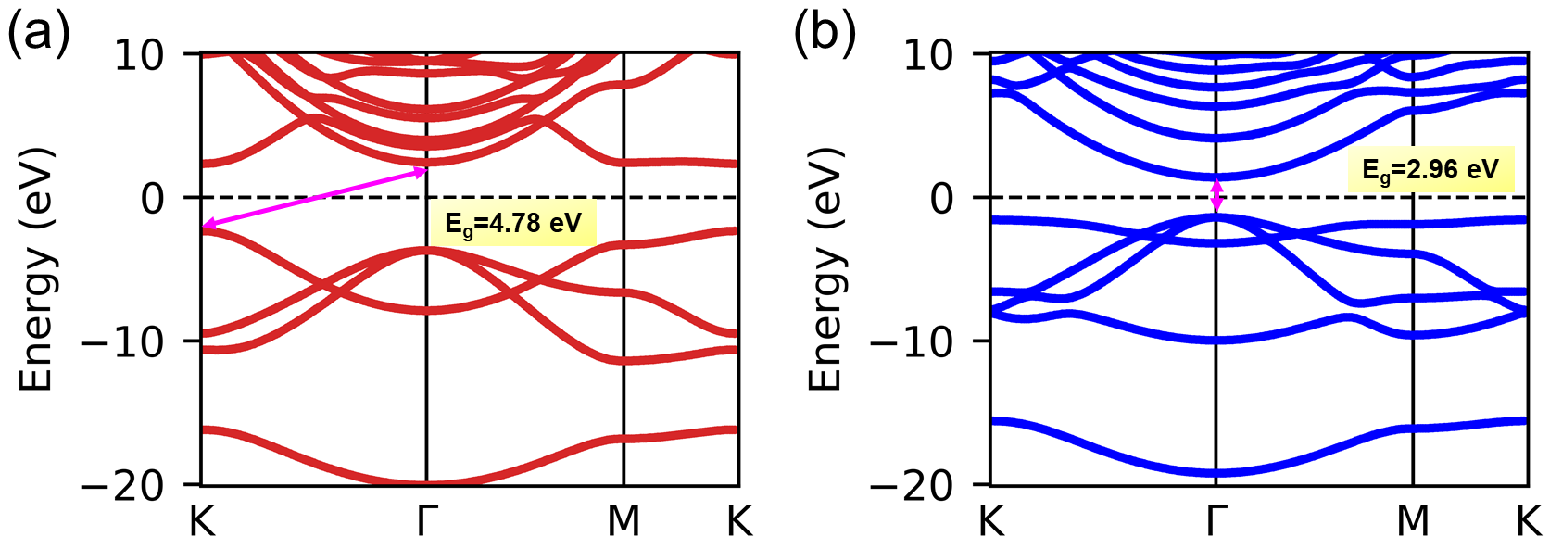}
\caption{Electronic band structures of monolayer (a) \textit{h}-BN and (b) H$_2$BN along the high-symmetry directions in the first Brillouin zone. The chemical potential is set to zero. In inset, the E$_g$ represents the energy gap between the valence band (VB) maximum and the conduction band (CB) minimum.}
\label{fig:S1}
\end{figure*} 
\indent {Fig.S2 shows the electronic density of states (DOS) of monolayer \textit {h}-BN (pristine) and fully hydrogenated \textit {h}-BN (H$_2$BN). While the highest occupied electronic states are of dominantly of N 2$p_z$ characters for monolayer \textit {h}-BN (Fig.S2a), the most of the DOS of H$_2$BN dominantly come from B 2$p_z$ and H1s orbitals and partly from N 2$p_x$ and 2$p_y$ (Fig.S2b) when summing contributions for all k-points (Note, however, that for the $\Gamma$ point-resolved DOS just below the chemical potential the major contributions arise from N 2$p_x$ and 2$p_y$ orbitals whereas minor come from B 2$p_z$ and H 1s orbitals. Contributions are reversed for the high-symmetry K-point-resolved DOS just below the chemical potential). The B 2$p_z$ and H 1 s states are strongly hybridized with almost similar density peak in the projected DOS spectra (Fig.S2b). The B 2$p_z$ state which dominantly appears just above the chemical potential (Fig.S2a), now disappears when \textit {h}-BN is fully hydrogenated (Fig.S2b). So, the hydrogenation results in the significant modifications in the electronic properties of \textit {h}-BN. Since the hydrogen electronic states have energy closer to the Fermi energy. The suitable doping of H$_2$BN by holes may bring these hydrogen electronic states to the Fermi level. The corresponding electrons then may interact with phonons, thereby contributing to the e-ph coupling.}
\begin{figure*}[h!]
\includegraphics[scale=1]{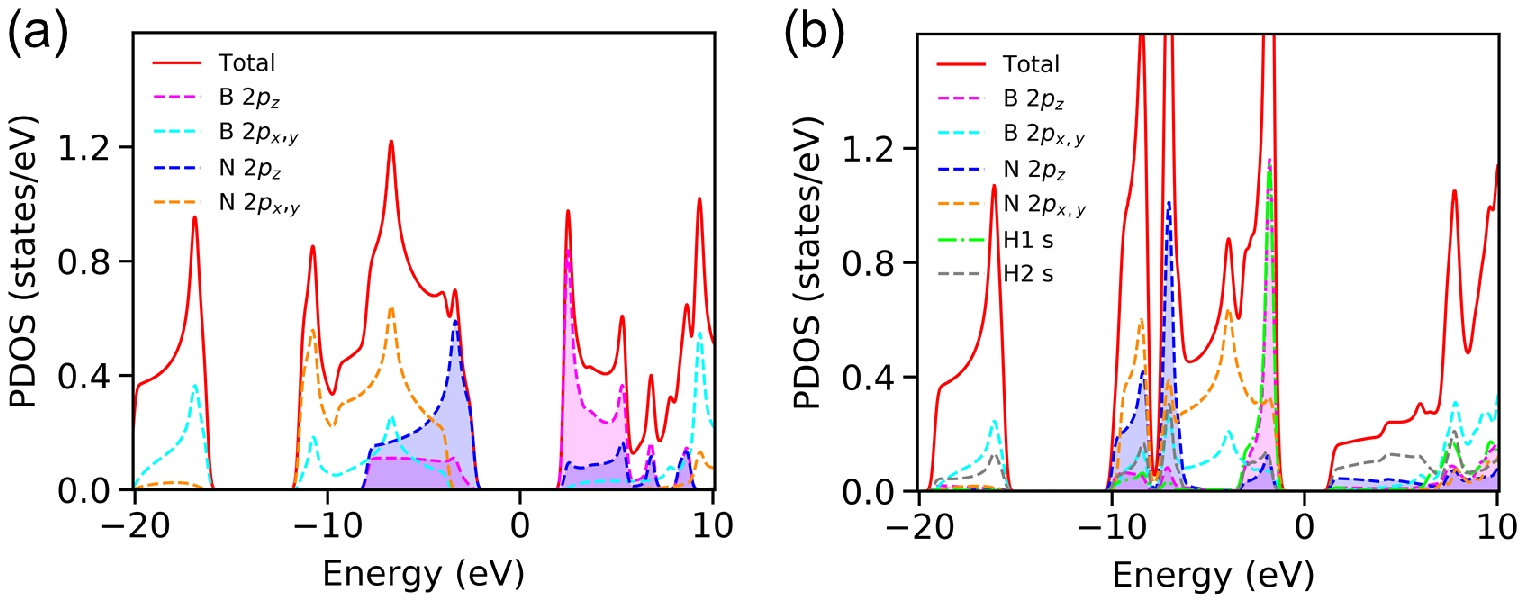}
\caption{Projected density of states (PDOS) of (a) monolayer \textit{h}-BN and (b) fully hydrogenated monolayer \textit{h}-BN (H$_2$BN). The chemical potential is set to zero. In a) the B 2$p_z$ state of monolayer \textit{h}-BN is partly unfilled just above the chemical potential (pink trace) where the density of B 2$p_z$ is relatively higher than that below the chemical potential. In b), for the case of H$_2$BN, both B 2$p_{z}$ and H1 s states dominantly contribute to the DOS just below the chemical potential, while the N and B 2$p_{x,y}$ and H2 s orbitals also contribute to the DOS. The contributions from 2$p_x$ and 2$p_y$ are added and shown as 2$p_{x,y}$. Here, H1 and H2 are bonded to B and N atoms, respectively.}
\label{fig:S2}
\end{figure*} 
\begin{figure*}[h!]
\includegraphics[scale=1]{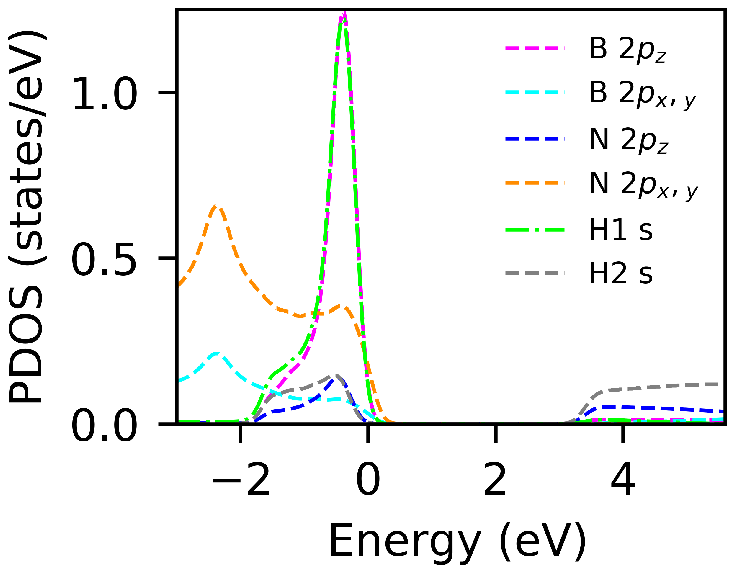}
\caption{Projected density of states (PDOS) of hole-doped H$_2$BN with a carrier density of $1.0\times 10^{14}$ holes cm$^{-2}$. Here, the PDOS is shown for the selected energy range from -3.0 eV to 5.6 eV. The Fermi level is set to zero. Within from $\sim$ -1.8 eV to 0.2 eV, B $2p_z$ (magenta trace) and H1 s (green) states are strongly hybridized. The N 2$p_{x,y}$ also have contributions to this hybridized state. The H1 atom which is bonded to B atom has similar PDOS near and at $E_F$ to that coming from B $2p_z$ orbital.} 
\label{fig:S3}
\end{figure*}\\
\indent {We now discuss the electronic DOS of the hole-doped H$_2$BN with the carrier density of $1.0\times 10^{14}$ holes cm$^{-2}$ obtained by projection onto the atomic orbitals of B, N, and H atoms. In Fig.S3, we show the DOS projected on B and N 2p orbitals and H s orbitals. The hole doping brings some electronic states at the Fermi level. These states mainly include the B $2p_z$, H 1s, and N 2$p_{x}$ and 2$p{_y}$ components (Note again that the resulting DOS is obtained by considering all k-points in the Brillouin zone. For the $\Gamma$ point-resolved DOS, the major contributions arise from N 2$p_x$ and 2$p_y$ orbitals.)}
\section {Vibrational properties of undoped and doped H$_2$BN}
\noindent {We now determine the dynamical stability of the undoped and hole-doped H$_2$BN systems. We evaluate the phonon frequency dispersion for both undoped and doped H$_2$BN without strain. In Fig.S4, we show the phonon frequency dispersion of the doped H$_2$BN with a carrier density of $1.0\times$ 10$^{14}$ holes cm$^{-2}$, and compare it with that of undoped H$_2$BN. The hole-doped H$_2$BN model exhibits a smooth softening of the optical and acoustic branches around the zone center with respect to the undoped H$_2$BN. The optical B-N stretching mode is softened when H$_2$BN is doped. The hole-doped H$_2$BN system with a suitable density of hole-carrier thus exhibits the Kohn anomaly in phonon dispersion. The other optical modes associated with H motion are also found to be softened. These results suggest that upon further doping these optical and acoustic modes may be further softened until the lattice becomes dynamically unstable.}
\begin{figure*}[h!]
\includegraphics[scale=1]{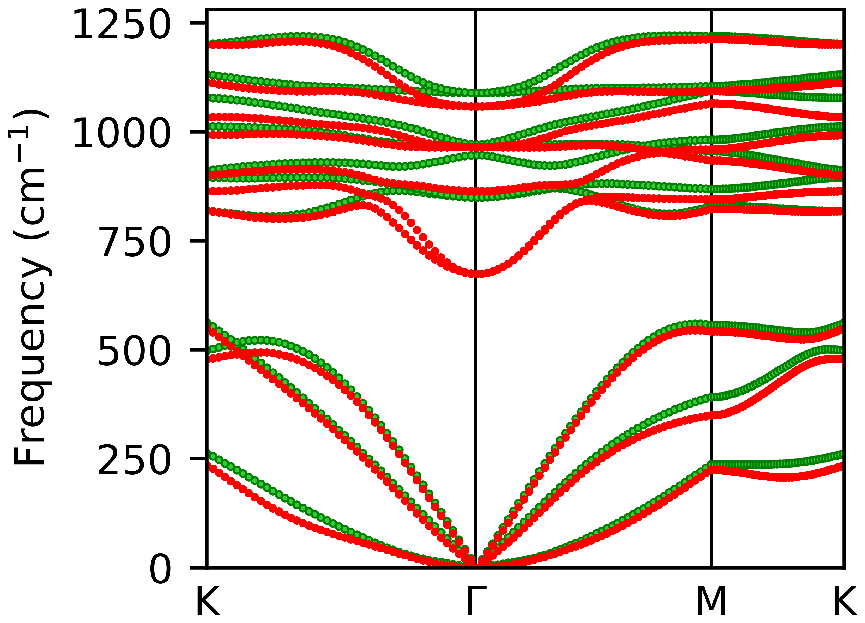}
\caption{Comparison of phonon dispersion of the undoped H$_2$BN (green trace) and $1.0\times$ 10$^{14}$ holes cm$^{-2}$ (red) for the selected frequency range up to 1280 cm$^{-1}$, showing the Kohn anomaly.}
\label{fig:S4}
\end{figure*}\\
\indent {In Table S1, we present the selected phonon frequencies calculated at the zone center of the Brillouin zone of undoped and hole-doped H$_2$BN systems, and the corresponding normal modes and symmetry of vibrations. The low-frequency optical modes which are related to H motions and the optical B-N modes are found to soften when H$_2$BN system is doped with carrier density of $0.3\times$ 10$^{14}$ holes cm$^{-2}$ at which in-plane B-N stretching mode at the $\Gamma$ point is softened by 67 cm$^{-1}$ (from 850 cm$^{-1}$ to 783 cm$^{-1}$). Similarly, an optical mode associated with in-plane H motion is softened by 16 cm$^{-1}$ (from 1089 cm$^{-1}$ to 1073 cm$^{-1}$). With further increase of holes to $1.0\times$ 10$^{14}$ holes cm$^{-2}$ and $1.5\times$ 10$^{14}$ holes cm$^{-2}$, these modes become more softer (see Table S1). The softening of the in-plane phonon modes may play an important role in enhancing the phonon-mediated superconductivity.}\\
\setlength{\tabcolsep}{10pt}
\begin{table*}[h!]
\renewcommand\thetable{S1} 
\caption{Comparison of the selected optical phonon frequencies (given in unit of cm$^{-1}$) at the center of Brillouin zone of undoped H$_2$BN and hole-doped H$_2$BN systems. For the doped H$_2$BN, the carrier densities (in unit of 10$^{14}$ holes cm$^{-2}$) are given in parenthesis. The symmetry and normal modes are also provided.}
\begin{tabular}{c c c c c c } 
\hline
Symmetry  &  Normal modes &  H$_2$BN & H$_2$BN & H$_2$BN & H$_2$BN\\ [0.5ex]
&&&(0.3)& (1.0) & (1.5)\\
\hline
E       &   In-plane H         &    1089  &  1073 & 1059 & 1056\\ [1ex]
E       &   In-plane H (90$\degree$ rotation)       &    1089  &  1073 & 1059 & 1056\\ [1ex]
E       &   In-plane H (30$\degree$ rotation)      &    971   &  970 & 965 & 956\\ [1ex]
E       &   In-plane H  (120$\degree$ rotation)      &    971   &  970 & 965 & 956\\ [1ex]
A$_1$   &   Out-of-plane B-N &    946  &  908 & 864 & 843\\ [1ex]
E'      &   B-N stretching    &    850   &  783 & 675 & 647\\ [1ex]
E'      &   B-N stretching (90$\degree$ rotation)    &    850   &  783 & 675 & 647\\ [1ex]
\hline
\label{S1}
\end{tabular}
\end{table*}
\begin{figure}[h!]
\includegraphics[scale=1]{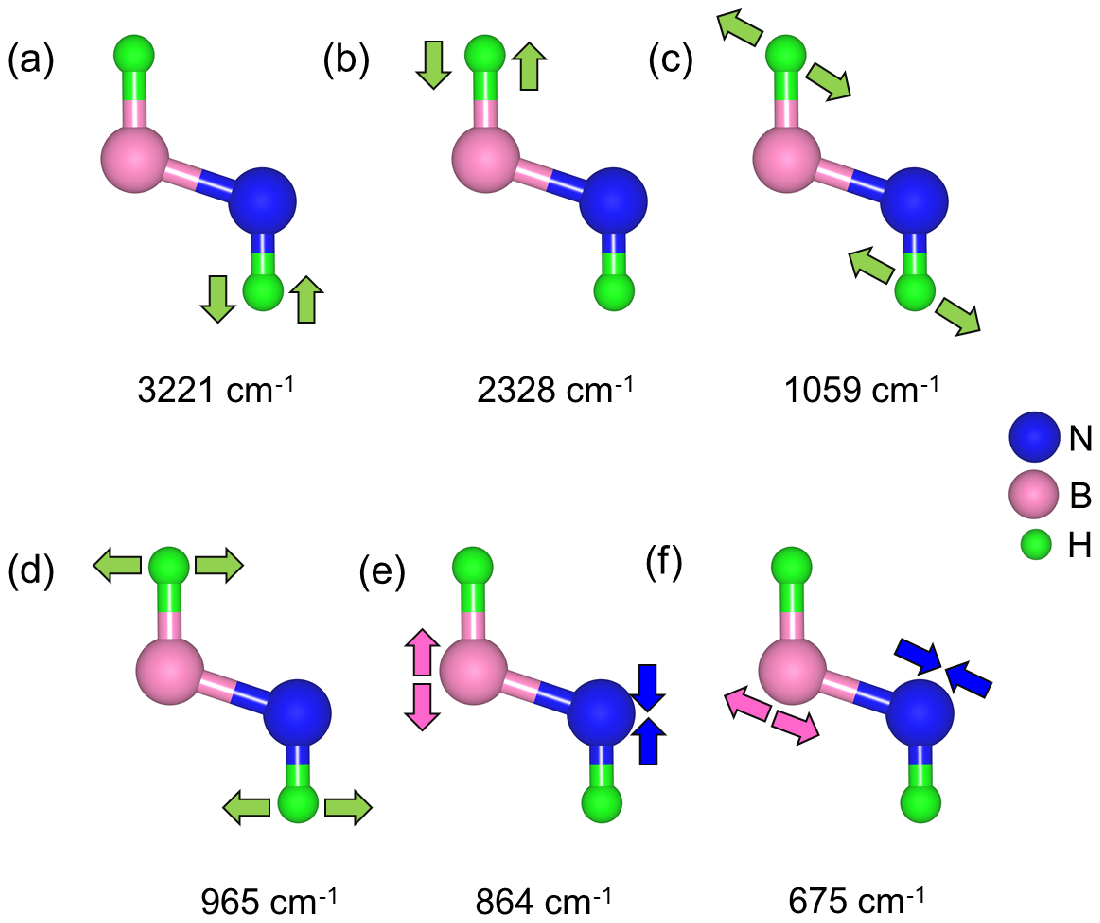}
\caption{Schematic representations of the selected optical phonon modes at the $\Gamma$ point of Brillouin zone of the hole-doped H$_2$BN with a hole carrier density of $1.0\times$ 10$^{14}$ holes cm$^{-2}$. The green, pink, and blue arrows indicate the motions of hydrogen, boron, and nitrogen atom, respectively.}
\label{fig:S5}
\end{figure}\\
\\
\indent {In Fig.S5, we show the selected optical phonon modes with corresponding frequencies that are calculated at the center of the Brillouin zone of the hole-doped H$_2$BN with the carrier density of $1.0\times$ 10$^{14}$ holes cm$^{-2}$. The high-frequency modes with frequencies of 3221 cm$^{-1}$ and 2328 cm$^{-1}$ are associated with the H motion nearly perpendicular to the lattice plane, but at other high symmetry points, the polarization of H-atom vibrations may not be perpendicular to the lattice plane. Our numerical results suggest that these modes of H atoms are little re-normalized or suppressed by the hole doping. At the $\Gamma$ point, the transverse optical mode with phonon frequency of 864 cm$^{-1}$ associated with the motion of B and N atoms is also nearly perpendicular to the plane. At the $\Gamma$ point, the in-plane B-N stretching mode with phonon frequency of 675 cm$^{-1}$ should contribute dominantly, followed by two modes with 1059 cm$^{-1}$ and 965 cm$^{-1}$ associated with the in-plane motion of H atoms, to the e-ph interaction.}
\begin{figure*}[h!]
\includegraphics[scale=1]{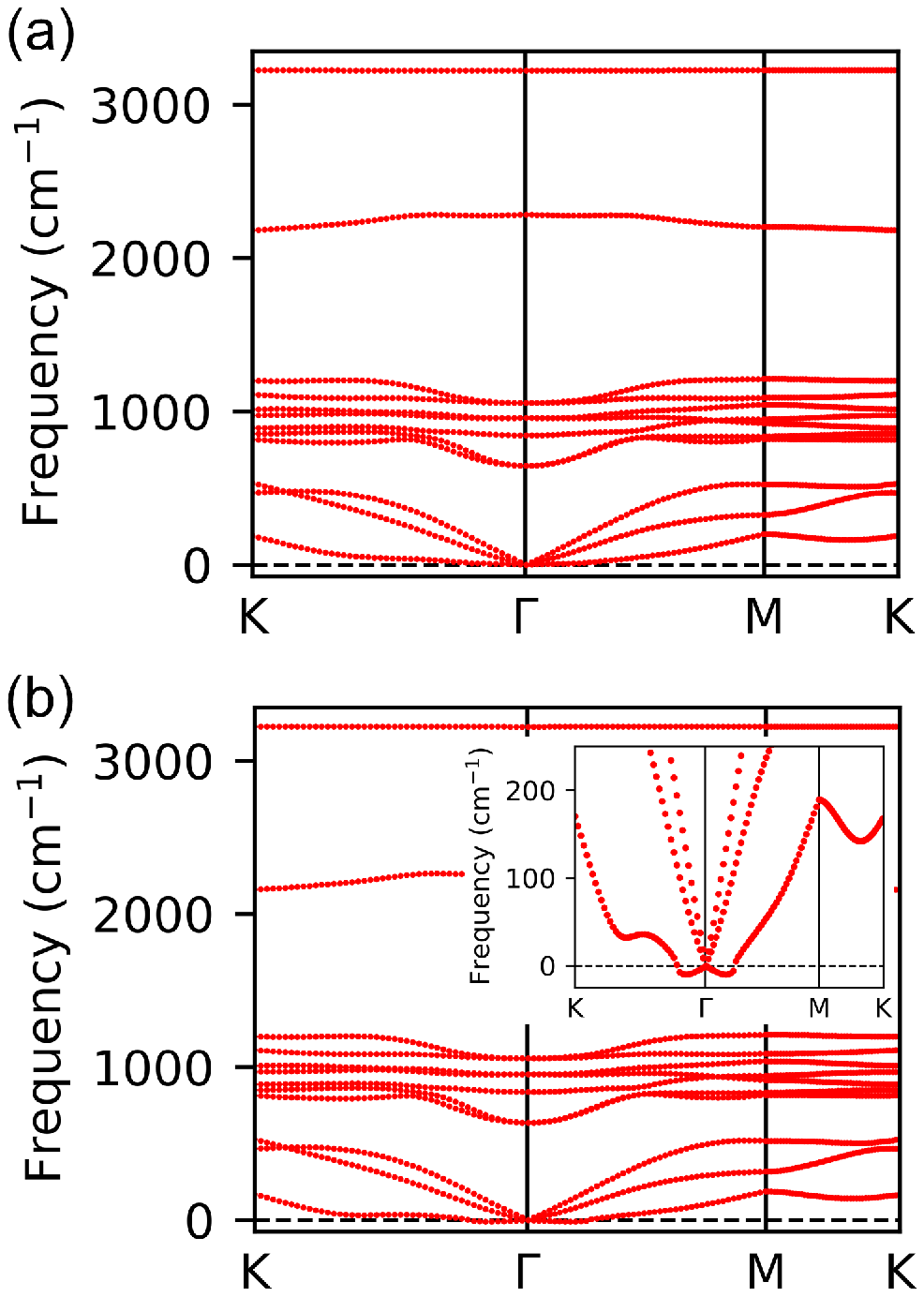}
\caption{Phonon frequency dispersion of the doped H$_2$BN with the carrier density of (a) $1.5\times$ 10$^{14}$ holes cm$^{-2}$, and (b) $1.6\times$ 10$^{14}$ holes cm$^{-2}$. For the latter hole carrier density, the doped H$_2$BN system has negative frequencies shown in inset.}
\label{fig:6}
\end{figure*}\\
\indent {We now examine the dynamical stability of doped H$_2$BN for higher hole carrier densities. We present the phonon frequency dispersion of doped H$_2$BN near and at the lattice instability. In Fig.S6, we present the phonon frequency dispersion of doped H$_2$BN with carrier densities of $1.5\times$ 10$^{14}$ holes cm$^{-2}$ and $1.6\times$ 10$^{14}$ holes cm$^{-2}$. The doped H$_2$BN system is found dynamically stable for the hole carrier density of $1.5\times$ 10$^{14}$ holes cm$^{-2}$ (Fig.S6a), but the system is not stable after we increase the carrier density to $1.6\times$ 10$^{14}$ holes cm$^{-2}$ (Fig.S6b). For the latter hole carrier density, the system has negative frequencies near the $\Gamma$ point (Fig.S6b, inset). Therefore, this could be the maximum hole carrier density that can be utilized for doping the unstrained H$_2$BN model.}\\
\section {Effect of strain on the vibrational properties of undoped and doped H$_2$BN}
\noindent {In Table S2, we show the selected phonon frequencies and modes calculated at the center of Brillouin zone of the hole-doped H$_2$BN under the biaxial tensile strain (BTS) of 6 \% up to the hole carrier density of $3.4\times$ 10$^{14}$ holes cm$^{-2}$. The optical modes associated with H in-plane motion, are found to be softer with frequencies of 985 cm$^{-1}$ and 886 cm$^{-1}$. Such softening of these H motion-related modes are expected to give strong e-ph interaction.}\\
\setlength{\tabcolsep}{10pt}
\begin{table*}[h!]
\renewcommand\thetable{S2} 
\caption{Comparison of selected optical phonon frequencies (given in unit of cm$^{-1}$) at the center of Brillouin zone of undoped H$_2$BN and hole-doped H$_2$BN systems with BTS of 6 \%. The carrier densities (in unit of 10$^{14}$ holes cm$^{-2}$) are given in parenthesis. The symmetry and normal modes are also provided.}
\begin{tabular}{ c c c c c c} 
\hline
Symmetry  &  Normal modes &  H$_2$BN & H$_2$BN & H$_2$BN & H$_2$BN\\ [0.5ex]
&&&(1.6)& (2.0) & (3.4)\\
\hline
E       &   In-plane H        &    1052  &  1013 & 997 & 985\\ [1ex]
E       &   In-plane H (90$\degree$ rotation)       &    1052  &  1013 & 997 & 985\\ [1ex]
E       &   In-plane H (30$\degree$ rotation)      &    953   &  935 & 912 & 886\\ [1ex]
E       &   In-plane H  (120$\degree$ rotation)      &    953   &  935 & 912 & 886\\ [1ex]
A$_1$   &   Out-of-plane B-N &    925  &  887 & 886 & 885\\ [1ex]
E'      &   B-N stretching    &    734   &  710 & 709 & 707\\ [1ex]
E'      &   B-N stretching (90$\degree$ rotation)    &    734   &  710 & 709 & 707\\ [1ex]
\hline
\label{S2}
\end{tabular}
\end{table*}
\begin{figure*}[h!]
\includegraphics[scale=1]{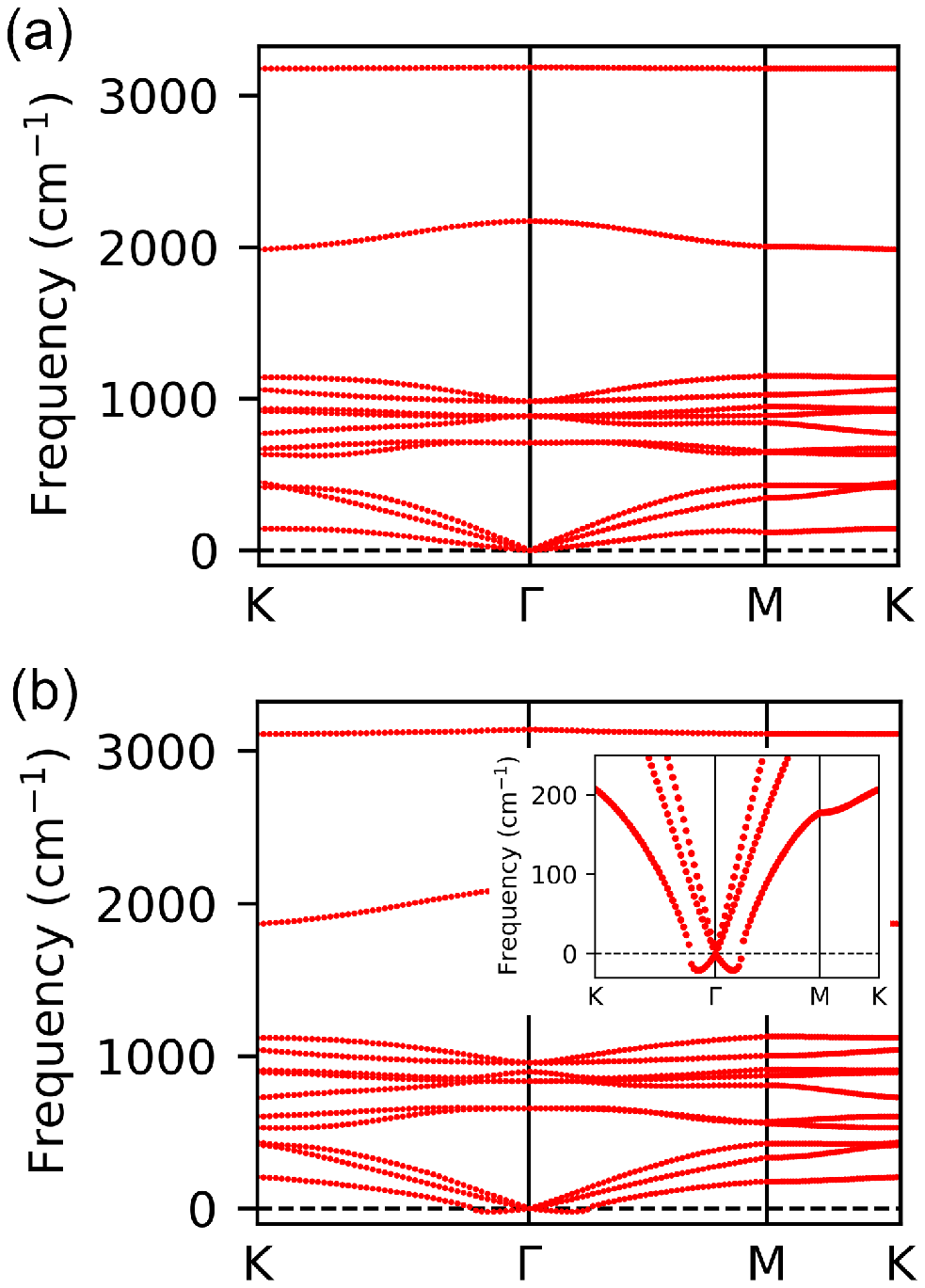}
\caption{Phonon frequency dispersion of the strained and doped H$_2$BN models under 6 \% BTS with the carrier densities of (a) $3.4\times$ 10$^{14}$ holes cm$^{-2}$, and (b) $3.5\times$ 10$^{14}$ holes cm$^{-2}$. For the latter hole carrier density, the doped H$_2$BN system has negative frequencies around the $\Gamma$ point, as shown in inset. Under the BTS of 6 \%, the critical hole density ($n_c$) that can be injected to H$_2$BN appears to be $\sim3.45\times$ 10$^{14}$ holes cm$^{-2}$.}
\label{fig:S7}
\end{figure*}\\
\indent{We determine the effect of BTS on the phonon frequency dispersion of doped H$_2$BN models. After applying a 6 \% BTS, we can inject more holes to the H$_2$BN model up to $\sim$ $3.4\times$ 10$^{14}$ holes cm$^{-2}$. The optical phonon frequencies at the center of Brillouin zone of undoped H$_2$BN model with BTS of 6 \% (Table S2) are found to be smaller than those at the zone-center of undoped H$_2$BN (Table S1). The dispersion features around the $\Gamma$ point in the Brillouin zone of the doped H$_2$BN with a carrier density of $3.4\times$ 10$^{14}$ holes cm$^{-2}$ under BTS of 6 \% (Fig.S7a) are found to be different from those of doped H$_2$BN (unstrained) with a carrier density of $1.5\times$ 10$^{14}$ holes cm$^{-2}$ (Fig.S6a). The further differences can also be understood by comparing phonon frequencies in Table S2 with those listed in Table S1.}\\
\indent{In Fig.S7, we compare the phonon frequency dispersion of strained and doped H$_2$BN with the carrier density of $3.4\times$ 10$^{14}$ holes cm$^{-2}$ and $3.5\times$ 10$^{14}$ holes cm$^{-2}$. For the former carrier density, the system is found stable (Fig.S7a). With further increasing of the hole carrier density up to $3.5\times$ 10$^{14}$ holes cm$^{-2}$, the doped H$_2$BN system exhibits the negative frequencies near the $\Gamma$ point (Fig.S7b inset), thus indicating the system to be dynamically unstable. As shown in Fig.S7a, the the H$_2$BN lattice with a BTS up to 6 \% is found to be dynamically stable up to this carrier density near the lattice instability. For the hole doping of $3.4 \times$ 10$^{14}$ holes cm$^{-2}$, both the optical phonon and acoustic branch related to the in-plane displacements of B and N atoms and the in-plane motion of H atoms may give rise to the strong e–ph coupling and therefore the phonon-mediated superconductivity.}
\newpage
\section {Details of estimating the superconducting transition temperature}
Figs.S8 and S9 show the variations of the superconducting transition temperature ($T_c$) as functions of the degauss parameter for phonons and smearing (or broadening) parameter for e-ph interaction. We firstly calculate $T_c$ as a function of smearing parameters at several different and finite degauss parameters until the results become convergent. The convergent result is shown in Fig.S8. The values of degauss in the range from 0.03 to 0.1 THz (or equivalently 1.0 to 3.3 cm$^{-1}$) may be sufficient to do smearing for phonon DOS. The variation in $T_c$ as a function of the double delta smearing parameter is shown in Fig.S9 with the choice of degauss of 0.1 THz. In the mathematical expressions for the electron DOS, there exist the $\delta$-functions. To simplify numerical calculations, the $\delta$-function is usually replaced by a Lorentz distribution function with a smearing parameter. The correct limit is to choose $\gamma_s$ as small as possible, and only in $\gamma_s$ $\rightarrow$0 limit, the Lorentzian function goes back to $\delta$-function.
\begin{figure*}[h!]
\includegraphics[scale=1]{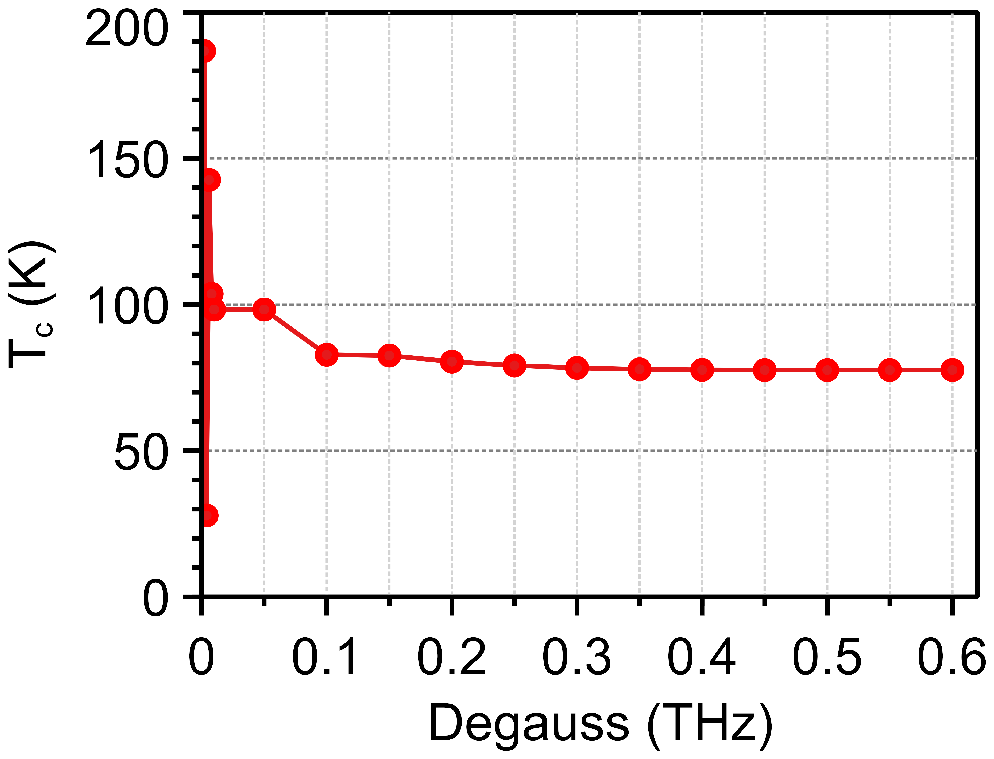}
\caption{Superconducting transition temperature for hole-doped H$_2$BN with BTS=6 \% and carrier density of $3.4\times$ 10$^{14}$ holes cm$^{-2}$ as a function of degauss for phonons.}
\label{fig:S8}
\end{figure*}
\begin{figure*}[h!]
\includegraphics[scale=1]{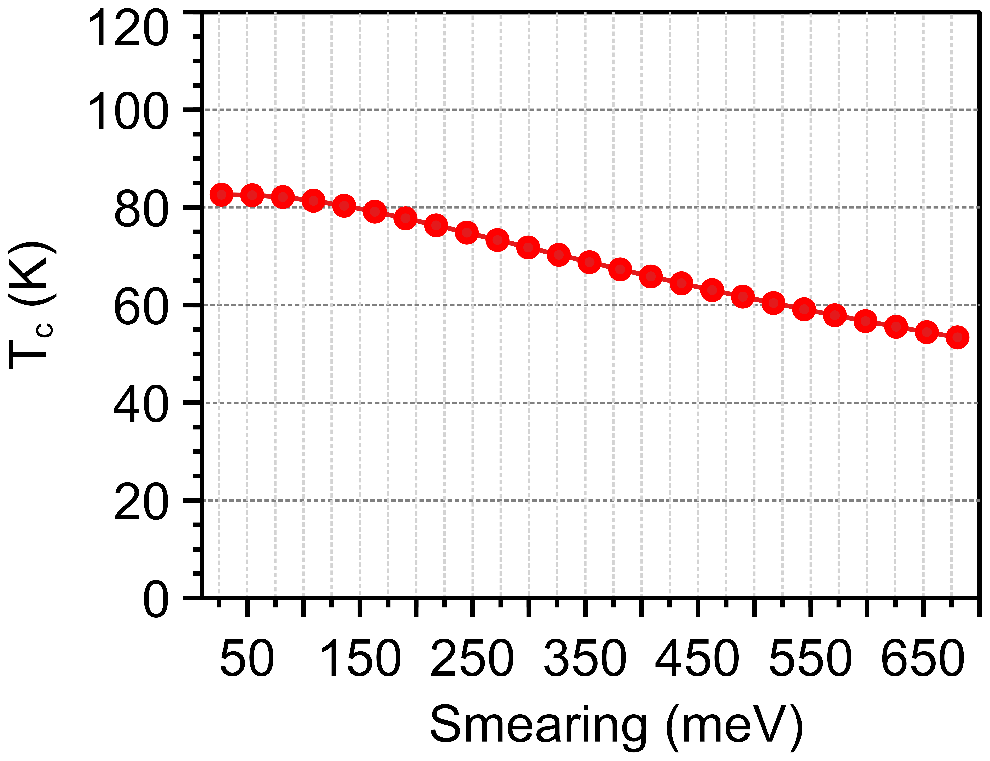}
\caption{Superconducting transition temperature for hole-doped H$_2$BN with BTS=6 \% and carrier density of $3.4\times$ 10$^{14}$ holes cm$^{-2}$ as a function of double delta smearing for e-ph interaction.}
\label{fig:S9}
\end{figure*}
\end{document}